\documentstyle[preprint,aps]{revtex}
\begin{document}
\draft
\preprint{\vbox{
\hbox{TRI-PP-95-55}
\hbox{UCRHEP-T150}
\hbox{hep-ph/9508338}
\hbox{August 1995}
}}
\title{The $Z \rightarrow b \bar b$ Excess and Top Decay}

\author{Ernest Ma}
\address{Department of Physics, University of California, Riverside, CA 92521}

\author{Daniel Ng}
\address{TRIUMF, 4004 Wesbrook Mall, Vancouver, B.C., V6T 2A3, Canada}

\maketitle

\begin{abstract}
The apparent excess of $Z \rightarrow b \bar b$ events at LEP may be an
indication of new physics beyond the standard model.  However, in either
the two-Higgs-doublet model or the minimal supersymmetric standard model,
any explanation would lead to an important new decay mode of the top quark
and suppresses the $t \rightarrow W b$ branching fraction, which goes
against what has been observed at the Tevatron.  In the two-Higgs-doublet
model, the branching fraction of $Z \rightarrow b \bar b$ + a light boson
which decays predominantly into $b \bar b$ would be at least of order
10$^{-4}$.
\end{abstract}
\newpage

\section{INTRODUCTION}
\label{sec:int}

With an accumulation of $8 \times 10^6$ $Z$ decays into hadrons and charged
leptons by the four LEP experiments at CERN by the end of 1993, measurements
of a large number of rates, branching fractions, and asymmetries have now
become even more precise\cite{lep}. The only apparent deviation by two
or more standard deviations from the prediction of the standard model is in
the ratio
\begin{equation}
R_b \equiv {{\Gamma (Z \rightarrow b \bar b)} \over {\Gamma (Z \rightarrow
{\rm hadrons})}}.
\end{equation}
Assuming $m_t = 175$ GeV and $m_H = 300$ GeV, the standard model predicts
that $R_b = 0.2158$, whereas LEP obtained $R_b = 0.2202 \pm 0.0020$ if the
similarly defined $R_c$ is assumed to be independent.  If the latter is fixed
at its standard-model value, then $R_b = 0.2192 \pm 0.0018$.  In either case,
the excess is about 2\% $\pm$ 1\%.  If this is taken seriously, physics
beyond the standard model is indicated.

In this paper we will examine two frequently studied extensions of the
standard model: the two-Higgs-doublet model (2HDM) and the minimal
supersymmetric standard model (MSSM).  We will assume that the only
significant deviation from the standard model is $R_b$, hence we will
take $R_b = 0.2192 \pm 0.0018$ as the experimental value and see how
these two extensions may be able to explain it.  We will concentrate on
obtaining $R_b > 0.2174$, {\it i.e.} within one standard deviation of the
experimental value, because otherwise the difference with the standard
model is insignificant and we might as well not bother with any possible
extension.

Whereas the contributions to $R_b$ from either the
2HDM\cite{denner,liu&ng,grant,park} or the MSSM\cite{barbieri,wells,garcia}
have been studied previously, we are concerned
here also with the effect of these new contributions on top decay.  Together
with the constraints from the oblique parameters\cite{peskin}, we find that
a large $R_b$ excess will always lead to an important new decay mode of the
top quark and suppresses the $t \rightarrow Wb$ branching fraction, which
goes against what has been observed at the Fermilab Tevatron\cite{tevatron}.
In the 2HDM, we also find that the branching fraction of $Z \rightarrow b
\bar b$ + a light boson which decays predominantly into $b \bar b$ will be
at least of order 10$^{-4}$\cite{ma&ng}.

\section{Two Higgs Doublets}
\label{sec:2hdm}

The simplest extension of the standard model is to have two Higgs doublets
instead of just one.  The relevance of this model to $R_b$ was studied in
detail already a few years ago\cite{denner}.  To establish notation, let
the two Higgs doublets be given by
\begin{equation}
\Phi_i = \left( \begin{array} {c} \phi_i^+ \\ \phi_i^0 \end{array} \right) =
\left( \begin{array} {c} \phi_i^+ \\ (v_i + \eta_i + i \chi_i)/\sqrt 2
\end{array} \right)\; .
\end{equation}
Let $\tan \beta \equiv v_2/v_1$, then
\begin{eqnarray}
h^+ &=& \phi_2^+ \cos \beta - \phi_1^+ \sin \beta \;, \\
A &=& \chi_2 \cos \beta - \chi_1 \sin \beta \; , \\
h_1 &=& \eta_1 \cos \alpha + \eta_2 \sin \alpha \; , \\
h_2 &=& \eta_2 \cos \alpha - \eta_1 \sin \alpha \; .
\end{eqnarray}

The corrections to the left- and right-handed $Z b \bar b$ vertex
induced by the charged Higgs boson $h^+$ and the neutral Higgs bosons $h_1$,
$h_2$ and $A$ are given by\cite{denner}
\begin{eqnarray}
\delta g_{L,R}^b(h^+) = \frac{\alpha}{4\pi s_W^2} \lambda_{L,R}^2
	~\rho^C_{L,R}(q^2,m_+^2,m_t^2) \; ,
\label{dgC}
\end{eqnarray}
and
\begin{eqnarray}
\delta g_{L,R}^b(h_{1,2},A) = \frac{\alpha}{4\pi s_W^2}
  \left(\frac{m_b\tan\beta}{2m_W}\right)^2 [
   \cos^2\alpha ~\rho_{L,R}^N(q^2,m_1^2,m_A^2) +
   \sin^2\alpha ~\rho_{L,R}^N(q^2,m_2^2,m_A^2) ] \; ,
\label{dgN}
\end{eqnarray}
where
\begin{eqnarray}
\rho_{L,R}^C(q^2,m^2,M^2) =&& \{-,+\}\textstyle{\frac{1}{2}}m_t^2
 C_0(q^2,m^2,M^2,M^2)+g_{L,R}^b~\rho_3(q^2,m^2,M^2,M^2)
\nonumber \\
&& +g^h \left[\rho_3(q^2,m^2,M^2,M^2)-\rho_4(q^2,m^2,m^2,M^2) \right] \; ,
\label{eq:rhoC}
\\
\rho_{L,R}^N(q^2,m^2,M^2)=&& \{-,+\}\rho_4(q^2,m^2,M^2,0)+g_{R,L}^b\left[
   \rho_3(q^2,m^2,0,0)+\rho_3(q^2,M^2,0,0) \right] \; .
\end{eqnarray}
In the above, $g_L^b = -\textstyle{\frac{1}{2}}+\textstyle{\frac{1}{3}}s_W^2$,
$g_R^b = \textstyle{\frac{1}{3}}s_W^2$ and $g^h=\textstyle{\frac{1}{2}}-s_W^2$.
We have also assumed that $\Phi_2$ couples to the up-type quarks and
$\Phi_1$ to the
down-type quarks, hence
$\lambda_L = \textstyle{m_t \over \sqrt{2} m_W} \cot\beta $, and
$\lambda_R = \textstyle{m_b \over \sqrt{2} m_W} \tan\beta $.
The masses of $h^+$, $h_1$,
$h_2$ and $A$ are denoted by $m_+$, $m_1$, $m_2$, and $m_A$ respectively.
The functions $C_0$ and $\rho_{3,4}$ are defined in Ref.~[2]; see also
Appendix \ref{app:a}.

For a heavy top quark, it is well-known that $\rho_{L,R}^C(q^2,m_+^2,m_t^2)
\simeq \{-,+\} (\textstyle{-\frac{1}{2}})$.  Hence $g_L^b {\rm{Re}}\{
\delta g_L^b(h^+)\}$ and $g_R^b {\rm{Re}}\{\delta g_R^b(h^+)\}$ are
negative, thereby decreasing the value of $R_b$.  This means that the
$\tan \beta < 1$ region can be ruled out\cite{liu&ng,grant,park}.
In this region, $t \to b h^+$ also becomes the dominant decay for the
top quark\cite{erler} unless it is kinematically not allowed.
In fact, any significant reduction of the $t \to Wb$ branching fraction
is in conflict with the Tevatron data\cite{tevatron} because the number of
top events observed is such that even if we assume $B(t \to W b) = 1$,
the deduced experimental $t \bar t$ production
cross section is already larger than expected\cite{qcd}.

If $\tan \beta$ is large, the contribution from the neutral Higgs bosons
becomes important.  In other words, Eq.~(\ref{dgN}) must be considered even
though it is suppressed by $(m_b/m_W)^2$.  Note that since $R_b$ is
proportional to $(g_L^b + \delta g_L^b)^2$ + $(g_R^b + \delta g_R^b)^2$,
$\rho^N_L$ is more important than $\rho^N_R$ because $g_L^b >>
g_R^b$\cite{liu&ng}.  Again because $g_R$ is small, $\rho^N_L$ is
dominated by $\rho_4(q^2,m^2,M^2,0)$ in Eq.~(10).  In order that
$\rho_4(q^2,m^2,M^2,0)$ be positive and not too small, both $m$ and $M$
must be light, namely $m^2, M^2 < q^2$.
In particular, for $\rho_4(m_Z^2,m^2,M^2,0) \ge 0.2$,
both $m$ and $M$ should be less than $65$ GeV.

It was shown already in Ref.~[2] that for $\tan \beta = 70 \simeq 2 m_t/m_b$,
the $R_b$ excess peaks at about 4\% near $m_A = m_1 \simeq 40$ GeV for
$\alpha = 0$.  However, since $Z \rightarrow A h_1$ is not observed,
$m_A + m_1 > m_Z$ is a necessary constraint.
We show in Fig.~\ref{fig1} the
contours in the $m_1 - m_A$ plane for $R_b = 0.2192$ and 0.2174.  It is
clear that relatively light scalar bosons are required if the $R_b$ excess
is to be explained.

For $A(h_1)$ lighter than $m_Z$ and having an enhanced coupling to $b \bar b$,
the decay $Z \rightarrow b \bar b + A (h_1)$ becomes nonnegligible\cite{dzz}.
As an illustration, we show in Fig.~\ref{fig2}
the branching fractions of these two
decays as functions of $m_A$ with the constraint $m_A + m_1 = m_Z + 10$
GeV so that a reasonable fit to the $R_b$ excess is obtained.  It is seen
that the sum of these two branching fractions is at least of order 10$^{-4}$.
Once produced, $A$ or $h_1$ decays predominantly into $b \bar b$ as well.
Hence this scenario for explaining $R_b$ can be tested at LEP if the
sensitivity for identifying one $b \bar b$ pair as coming from $A$ or $h_1$
in $b \bar b b \bar b$ final states can be pushed down below $10^{-4}$.

Since $b_L$ is involved in any enhanced coupling to light particles in
explaining the $R_b$ excess, its doublet partner $t_L$ must necessarily
have the same enhanced coupling to related particles.  In the 2HDM, we
must have an enhanced $\bar t b h^+$ coupling.  Therefore, unless
$m_+ > m_t - m_b$, the branching fraction of $t \to b h^+$ will be
important.  As a result, the standard $t \to W b$ branching fraction will be
seriously degraded.  We show this in Fig.~\ref{fig3} as a function of $m_+$.
Large values of $m_+$ are disfavored in this scenario because the
splitting with $A$ and $h_1$ would result in a large contribution to the
oblique parameter $T$, resulting in the constraint $m_+ \le 150$
GeV\cite{grant}.

\section{Supersymmetric Higgs Sector}

In the minimal supersymmetric standard model (MSSM), the two Higgs doublets
have exactly the same gauge and Yukawa couplings as in the 2HDM we discussed
in the previous section.  In addition, the quartic scalar couplings of the
MSSM are determined by the gauge couplings and there are only three arbitrary
mass terms: $\Phi_1^\dagger \Phi_1$, $\Phi_2^\dagger \Phi_2$, and
$\Phi_1^\dagger \Phi_2 + \Phi_2^\dagger \Phi_1$.  Hence we need only two extra
parameters, usually taken to be $\tan \beta$ and $m_A$, to specify the entire
Higgs sector, subject of course to radiative corrections\cite{radhiggs}.
However, these corrections are only significant for small $\tan \beta$
and since we need a large $\tan \beta$ to explain $R_b$, we will use the
simpler tree-level expressions in our numerical analysis.  Combining both
the charged and neutral Higgs contributions, we plot in Fig.~\ref{fig4} the
$R_b = 0.2192$ and 0.2174 contours in the $m_A-\tan\beta$ plane, with
the constraint $m_A + m_1 > m_Z$.
We plot also the contours for $B(t \to W b) = 0.85$ and 0.7, which
correspond to reductions of 28\% and 51\% of the top
signals at Fermilab respectively.  It is abundantly clear that the Higgs
sector of the MSSM is not compatible with both a large $R_b$ and a small
$B(t \to b h^+)$.  In the 2HDM, $m_A$ and $m_+$ are independent parameters,
whereas in the MSSM, there is the well-known sum rule $m_+^2 = m_A^2 + m_W^2$.
Hence an approximate custodial symmetry exists in the MSSM to keep the
contribution to $T$ small, but at the same time $m_+ < m_t - m_b$ is
inevitable if $m_A$ is assumed to be small enough to obtain a large $R_b$
excess.  In Fig.~\ref{fig5} we plot the minimum branching fraction
$B(Z \to b \overline{b} A + b \overline{b} h_1)$ for a given lower bound
of $R_b$.  This minimum is obtained by varying $\tan\beta$ with a fixed
value of $m_A$.  As in Fig.~2,
we see that this branching fraction is at least of order $10^{-4}$.

\section{Charginos and Neutralinos}

In addition to the Higgs contributions, there are chargino ($\chi$) and
neutralino ($\cal N$)
contributions to $R_b$ in the MSSM.  They have
been studied previously\cite{barbieri,wells,garcia}, and it is known that
the particles in the loops
have to be light in order to obtain large contributions.
The parameters involved here are $\tan\beta,~\mu,~M_2,
{}~m_{{\tilde t}_{1,2}},~m_{{\tilde b}_{1,2}},
{}~\theta^{\tilde t}$, and $\theta^{\tilde b}$,
where the supergravity condition $M_1\sim 0.5 M_2$ at the
electroweak scale has been assumed.
The scalar mixing angles $\theta^{\tilde t}$ and $\theta^{\tilde b}$
as well as others are defined in Appendix \ref{app:b}.

Since scalar quarks have not been observed at LEP,
their masses must be greater than half of the center-of-mass energy, namely
$m_{\tilde t_1,\tilde b_1} \ge \textstyle{1\over2}m_Z$. On the other hand,
the lightest neutralino (${\cal N}_1$) is always lighter than the
lightest chargino ($\chi_1$), thus the condition $m_{\chi_1} \ge
\textstyle{1\over2}m_Z$ is not enough by itself.
Let us define the conservative constraints from the invisible width and
total width of $Z$
as $\delta\Gamma_{\rm inv} \equiv \Gamma_{\rm inv}({\rm expt})|_{\rm max}
-\Gamma_{\rm inv}({\rm SM})|_{\rm min}$ and $\delta \Gamma_Z \equiv \Gamma_Z
({\rm expt})|_{\rm max}$ - $\Gamma_Z ({\rm SM})|_{\rm min}$, where SM denotes
the standard model.

{}From the updated LEP data \cite{lep}, we obtain
\begin{eqnarray}
\Gamma(Z\to {\cal N}_1{\cal N}_1)&\le&\delta\Gamma_{\rm inv}=7.6~{\rm MeV}
\; ,\\
\Gamma(Z\to {\cal N}_i{\cal N}_j)&\le&\delta\Gamma_Z = 23~{\rm MeV} \; .
\label{constraints}
\end{eqnarray}
These constraints must be included for the analysis in order to provide
consistent results.

The corrections to the left- and right-handed $Z b \overline{b}$ vertex
induced by the charginos ($\chi$) and neutralinos (${\cal N}$) are given by
\begin{equation}
\delta g_{L,R}^b(\chi,{\cal N}) =
	\frac{\alpha}{4\pi s_W^2} F_{L,R}(\chi,{\cal N}) \ ,
\end{equation}
where
\begin{eqnarray}
\label{susyFL}
F_{L,R}(\chi) &=& {\Lambda_{jk}^{L,R}}^\ast~S^t_{ji}~\Lambda_{ik}^{L,R}~
        \rho_4(q^2,m^2_{{\tilde t}_i},m^2_{{\tilde t}_j},m^2_{\chi_k})
        -\Lambda_{ki}^{L,R}~{\cal O}^{L,R}_{ij}~{\Lambda_{kj}^{L,R}}^\ast~
        \rho_3(q^2,m^2_{{\tilde t}_k},m^2_{\chi_i},m^2_{\chi_j})
\nonumber \\
        &&+\Lambda_{ki}^{L,R}~({\cal O}^{R,L}_{ij}-{\cal O}^{L,R}_{ij})~
  		{\Lambda_{kj}^{L,R}}^\ast~m_{\chi_i}~m_{\chi_j}~
	C_0(q^2,m^2_{{\tilde t}_k},m^2_{\chi_i},m^2_{\chi_j}) \; ,\\
\label{susyFR}
F_{L,R}({\cal N}) &=& {\lambda_{jk}^{L,R}}^\ast~S^b_{ji}~\lambda_{ik}^{L,R}~
       \rho_4(q^2,m^2_{{\tilde b}_i},m^2_{{\tilde b}_j},m^2_{{\cal N}_k})
        -\lambda_{ki}^{L,R}~{\cal Q}^{L,R}_{ij}~{\lambda_{kj}^{L,R}}^\ast~
        \rho_3(q^2,m^2_{{\tilde b}_k},m^2_{{\cal N}_i},m^2_{{\cal N}_j})
\nonumber \\
        &&+\lambda_{ki}^{L,R}~({\cal Q}^{R,L}_{ij}-{\cal Q}^{L,R}_{ij})~
                {\lambda_{kj}^{L,R}}^\ast~m_{{\cal N}_i}~m_{{\cal N}_j}~
        C_0(q^2,m^2_{{\tilde b}_k},m^2_{{\cal N}_i},m^2_{{\cal N}_j}) \; ,
\end{eqnarray}
and there is an implicit sum over all repeated indices.  See Appendix B for
the definitions of the various quantities in the above.

For $\tan\beta < 20$ or $m_{{\tilde b}_{1,2}} > m_Z$, the chargino
contribution is the most important.  From Eq.(\ref{susyFL}), we see that
the contribution is the largest when the lighter scalar quark
${\tilde t}_1$ is mostly ${\tilde t}_R$, namely $\theta^{\tilde t}_{11}=0$.
We follow the usual strategy \cite{wells} of finding the maximally allowed
$R_b$ for a given $\tan\beta$ and $m_{\chi_1}$.  Here we impose also the
LEP constraints given by Eq.(\ref{constraints}).  In Fig. \ref{fig6}, we
plot the maximally allowed $R_b$ as a function of $\tan\beta$ for
$m_{\chi_1}=60$ GeV and $m_{{\tilde t}_1}=\textstyle{1\over2}m_Z$.
We see that $R_b > 0.217$ can be obtained.  However, top decay into
$\tilde t_1$ and a neutralino is now possible and the
corresponding branching fraction $B(t \to Wb)$ shows clearly that this
solution would conflict with the Tevatron data \cite{tevatron,qcd}.

If $\tan\beta$ is large, the neutralino contributions become
important because the $b$ quark coupling to the higgsino is
proportional to $1/\cos \beta$.  Here we would like to
point out that our $\tilde{b_i} \overline{{\cal N}_j^C} ^\ast b$
couplings given in
Eqs.~(\ref{lambdaL}) and (\ref{lambdaR}) are different from those given in
Ref.~\cite{wells} but agree with Ref.~\cite{gunion&haber}.
For simplicity and with little loss of generality, we assume that
$m_{{\tilde t}_1}=m_{{\tilde b}_1}=60$ GeV and
$m_{{\tilde t}_2}=m_{{\tilde b}_2}=250$ GeV.  We vary the scalar quark
mixing angles $\theta^{\tilde t}$ and $\theta^{\tilde b}$,
such that $R_b$ is maximum
within the allowed LEP constraints given by Eq.(\ref{constraints})
and $m_{\chi_1} > \textstyle{1\over 2}m_Z$.
This is the most optimistic scenario; we cannot
achieve a large enough $R_b$ otherwise.  Taking $\tan\beta=70$,
the contours of $R_b = 0.2174$ and $0.2192$ are plotted in Fig. \ref{fig7}.
Again we plot the $t\to Wb$ branching fraction.  We find only
very narrow regions where $B(t\to Wb) > 0.7$ and $R_b > 0.2174$.
Hence future experiments on top decay will play a decisive role
to verify or rule out this scenario\cite{misc}.

The dominant contribution to the oblique parameter $T$ comes from the scalar
quarks.  We have checked that $T \sim 0.4$ in the narrow regions, which is in
mild conflict with the recent global fit $T=-0.67\pm0.92$ \cite{burgess}.
Nevertheless, Fig.~\ref{fig7} represents the most optimistic scenario.
In Fig.~\ref{fig8}, we consider a more restrictive case with
$\theta^{\tilde t} = \theta^{\tilde b}$ so that $T = 0$.
As a result, the narrow regions shrink as expected.

\section{conclusion}

The 2HDM \cite{denner,liu&ng,grant,park} and the
MSSM\cite{barbieri,wells,garcia} have each been suggested to explain the
$R_b$ excess at LEP.  Here we consider also the effects of these new
contributions on top quark decay.  In the 2HDM, large $\tan\beta$ and
light neutral scalars are necessary to increase $R_b$ to within one
standard deviation of the experimental value.  The corresponding
$\bar t b h^+$ coupling then allows top decay into $b$ and $h^+$ unless
it is kinematically not allowed.  However, $m_+ > 150$ GeV would be in
conflict with the constraint of the oblique parameter $T$.  The same
interactions which allow a large $R_b$ also allow the decays
$Z\to b\bar{b}A(h_1)$.  We show in Fig.~2 that the branching fraction
$B(Z\to b\bar{b}A+b\bar{b} h_1)$ is at least of order $10^{-4}$ which
can be tested at future LEP experiments.

In the supersymmetric Higgs sector, there are only two independent
unknown parameters: $\tan\beta$ and $m_A$.  Specifically, because of the
sum rule $m_+^2 = m_A^2 + m_W^2$, top decay is always possible for a small
enough $m_A$ to account for the $R_b$ excess.  We see in Fig.~\ref{fig4} that
the region which allows $R_b$ to be large does indeed conflict with top
decay.  In addition, $B(Z\to b\bar{b}A+b\bar{b} h_1)$ is also at least of
order $10^{-4}$ as shown in Fig.~\ref{fig5}.

Because of the scalar-quark mixing angles, the chargino and neutralino
contributions to $R_b$ could each be either positive or negative.
Here we consider the
most optimistic scenario that the mixing angles are chosen to maximize the
$R_b$ value.  The chargino contribution, as opposed to the charged Higgs
and $W$ contributions, can increase $R_b$ above
$0.2174$ if ${\tilde t}_1$ and $\chi_1$ are light enough.  Since
${\cal N}_1$ is always lighter than $\chi_1$, we impose also the LEP
constraints given by Eq.~(\ref{constraints}).  Again this new
contribution gives rise to a new channel for top decay which reduces the
$t \to Wb$ branching fraction significantly.  For large $\tan\beta$, we
consider both chargino and neutralino contributions.  We find that there
are only very narrow regions, as shown in Figs.~\ref{fig7} and 8,
in which both large $R_b$ and $B(t\to Wb)$ are
compatible.  Future experiments on top decay would verify or rule out
this scenario.

{\it Note added.} After the completion of this paper, additional data from
LEP have been reported.\cite{renton}  For $R_c$ fixed at its standard-model
value, the latest experimental value of $R_b$ is $0.2205 \pm 0.0016$.
This means that our assertion that $R_b$ conflicts with top decay in the
2HDM and the MSSM becomes even stronger.

\vspace{0.5in}

\begin{center} {ACKNOWLEDGEMENT\\}
\end{center}

This work was supported in part by the U.~S.~Department of Energy under
Grant No. DE-FG03-94ER40837 and by the Natural Science and Engineering
Research Council of Canada.

\newpage

\begin{appendix}
\section{Definition of $\rho_3$}
\label{app:a}

The general form for $\rho_3$ is given by
\begin{eqnarray}
\rho_3(q^2,m^2,M_1^2,M_2^2)=
  &&[q^2(C_{22}-C_{23})+2C_{24}-M_1M_2C_0](q^2,m^2,M_1^2,M_2^2)\nonumber\\
  &&+{1\over2}[B_1(0,M_1^2,m^2)+B_1(0,M_2^2,m^2)]-{1\over2} .
\label{rho3}
\end{eqnarray}
For $M_1=M_2$, $\rho_3$ is identical to that defined in Ref.~\cite{denner}.
This generalized definition is useful in the supersymmetric case.

\section{Masses and Mixing in the MSSM}
\label{app:b}

In this appendix, we present the relevant couplings for the chargino and
neutralino contributions to the $Z b \overline{b}$ vertex.  We allow general
scalar top mixing, namely $\tilde{t}_L = \theta^{\tilde t}_{1i}~\tilde{t}_i$
and $\tilde{t}_R = \theta^{\tilde t}_{2i}~\tilde{t}_i$ with a similar
definition for the scalar bottom quarks.  The
$Z \tilde{t}_i^\ast \tilde{t}_j$ and
$Z \tilde{b}_i^\ast \tilde{b}_j$ vertices are given by
$\textstyle{g \over \cos\theta_W}~S^{t,b}_{ij}(p_i-p_j)_\mu$, where
\begin{eqnarray}
S^t_{ij}&=&{1\over2}{\theta^{\tilde t}_{1i}}^\ast\theta^{\tilde t}_{1j}-
	{2\over3}\sin^2\theta_W\delta_{ij}\; , \\
S^b_{ij}&=&-{1\over2}{\theta^{\tilde b}_{1i}}^\ast \theta^{\tilde b}_{1j}+
        {1\over3}\sin^2\theta_W\delta_{ij} \; .
\end{eqnarray}
The $p$'s are defined as out-going momenta.

The  chargino and neutralino mass matrices are given by
\begin{equation}
{\cal M}_\chi = \pmatrix{M_2 & \sqrt{2}m_W\cos\beta \cr
         \sqrt{2}m_W\sin\beta & \mu} \; ,
\end{equation}
which links $(i\tilde{W}^-,\tilde{h_1^-})^T$
to $(i\tilde{W}^+,\tilde{h_2^+})$, and
\begin{equation}
{\cal M}_{\cal N} =
\pmatrix{M_1 & 0 & -m_Z\sin\theta_W\cos\beta & m_Z\sin\theta_W\sin\beta \cr
      0 & M_2 & m_Z\cos\theta_W\cos\beta & -m_Z\cos\theta_W\sin\beta \cr
      -m_Z\sin\theta_W\cos\beta & m_Z\cos\theta_W\cos\beta & 0 & -\mu \cr
      m_Z\sin\theta_W\sin\beta & -m_Z\cos\theta_W\sin\beta & -\mu & 0\cr} \; ,
\end{equation}
in the $(i\tilde{B},i\tilde{W_3},\tilde{h_1^0},\tilde{h_2^0})$ basis.
The mass eigenstates, $\chi_i$ and ${\cal N}_i$, are related to
these basis states by the following transformations:
\begin{equation}
(i\tilde{W}^+,\tilde{h_2^+})^T = V_{ij} \chi_j \; ;\quad
(i\tilde{W}^-,\tilde{h_1^-})^T = U_{ij} \chi_j^C \; ,
\end{equation}
and
\begin{equation}
(i\tilde{B},i\tilde{W_3},\tilde{h_1^0},\tilde{h_2^0})^T = N_{ij}{\cal
N}_j \ .
\end{equation}
Thus, $M_\chi$ and $M_{\cal N}$ are diagonalized by
\begin{equation}
V^T~{\cal M}_\chi~U~=~m_{\chi_i}\delta_{ij} \; ,
\end{equation}
and
\begin{equation}
N^T~{\cal M}_{\cal N}~N~=~m_{{\cal N}_i}\delta_{ij} \; .
\end{equation}

Let us first consider the vertices for loops involving charginos
$\chi$.  The $Z \overline{\chi_i}\chi_j$ vertex is given by
$\textstyle{g \over \cos\theta_W}\gamma_\mu
[{\cal O}^L_{ij}~\textstyle{1-\gamma_5 \over 2}+
{\cal O}^R_{ij}~\textstyle{1+\gamma_5 \over 2}]$ with
\begin{eqnarray}
{\cal O}^L_{ij} &=& \cos^2\theta_W~\delta_{ij}
	-\textstyle{1\over2}V_{2i}^\ast~V_{2j} \; ,\\
{\cal O}^R_{ij} &=& \cos^2\theta_W~\delta_{ij}
        -\textstyle{1\over2}U_{2i}~U_{2j}^\ast \; ,
\end{eqnarray}
and the $\tilde{t_i}^\ast \overline{\chi_j^C} b$ vertex is given by
$g[\Lambda^L_{ij}~\textstyle{1-\gamma_5 \over 2}+
\Lambda^R_{ij}~\textstyle{1+\gamma_5 \over 2}]$ with
\begin{eqnarray}
\Lambda^L_{ij}&=&-V_{1i}~{\theta^{\tilde t}_{1j}}^\ast+
  {m_t\over\sqrt{2}m_W\sin\beta}~V_{2i}~{\theta^{\tilde t}_{2j}}^\ast \; , \\
\Lambda^R_{ij}&=&{m_b\over\sqrt{2}m_W\cos\beta}~
 U_{2i}^\ast~{\theta^{\tilde t}_{1j}}^\ast \; .
\end{eqnarray}
For the vertices involving neutralino loops,
$Z \overline{{\cal N}_i} {\cal N}_j$ vertex is given by
$\textstyle{g \over \cos\theta_W}\gamma_\mu
[{\cal Q}^L_{ij}~\textstyle{1-\gamma_5 \over 2}+
{\cal Q}^R_{ij}~\textstyle{1+\gamma_5 \over 2}]$ with
\begin{eqnarray}
{\cal Q}^L_{ij} &=& \textstyle{1\over2}(N_{3i}^\ast~N_{3j}-N_{4i}^\ast~N_{4j})
\; ,\\
{\cal Q}^R_{ij} &=& \textstyle{1\over2}(N_{4i}~N_{4j}^\ast-N_{3i}~N_{3j}^\ast)
\; ,
\end{eqnarray}
and the $\tilde{b_i}^\ast \overline{{\cal N}_j^C} b$ vertex is given by
$g[\lambda^L_{ij}~\textstyle{1-\gamma_5 \over 2}+
\lambda^R_{ij}~\textstyle{1+\gamma_5 \over 2}]$ with
\begin{eqnarray}
\label{lambdaL}
\lambda^L_{ij}&=&-{1\over3\sqrt{2}}~\tan\theta_W~N_{1i}~
   {\theta^{\tilde b}_{1j}}^\ast+
   {1\over\sqrt{2}}N_{2i}~{\theta^{\tilde b}_{1j}}^\ast-
   {m_b\over\sqrt{2}m_W\cos\beta}~N_{3i}~{\theta^{\tilde b}_{2j}}^\ast \; , \\
\label{lambdaR}
\lambda^R_{ij}&=&-{\sqrt{2}\over3}~\tan\theta_W~N_{1i}^\ast~
    {\theta^{\tilde b}_{2j}}^\ast-
 {m_b\over\sqrt{2}m_W\cos\beta}~N_{3i}^\ast~{\theta^{\tilde b}_{1j}}^\ast  \; .
\end{eqnarray}
Note that the relative signs of the $m_b$ terms in Eqs.~(\ref{lambdaL}) and
(\ref{lambdaR}) are in agreement with Ref.\cite{gunion&haber}, but differ from
Ref.\cite{wells}.

\end{appendix}

\begin{figure}

\caption{$R_b=0.2192$ (solid) and $0.2174$ (dashed)
contours in the $m_1 - m_A$ plane for $\alpha=0$ and $\tan \beta = 70$.
The straight line corresponds to $m_A + m_1 = M_Z$.  We have also assumed
$m_+ = m_2 = 175$ GeV.}
\label{fig1}
\end{figure}

\begin{figure}
\caption{The branching fractions, $B(Z \rightarrow b \bar{b} A)$ (dashed)
and $B(Z \rightarrow b \bar{b} h_1)$ (dotted) and their sum (solid), as
functions of $m_A$ where we take $m_A + m_1 = M_Z + 10$ GeV,
$\tan \beta = 70$, $\alpha = 0$, and $m_+ = m_2 = 175$ GeV.}
\label{fig2}
\end{figure}

\begin{figure}
\caption{The branching fraction $B(t \rightarrow Wb)$ as a function of
$m_+$ for $\tan \beta = 70$ (solid), 50 (dashed), and 20 (dotted).}
\label{fig3}
\end{figure}

\begin{figure}
\caption{$R_b=0.2192$ (heavy), $0.2174$ (solid) and
$B(t \rightarrow W b) = 0.85$ (dashed), $0.7$ (dotted) contours in the
$m_A-\tan\beta$ plane.  We also plot the constraint $m_A+m_1 > m_Z$
(dash-dotted).}
\label{fig4}
\end{figure}

\begin{figure}
\caption{The minimum branching fraction
$B(Z \rightarrow b \bar{b} A~+~b \bar{b} h_1)$ for a given lower bound
of $R_b$, where we take $m_A = 45$ GeV (solid) and $m_A = 50$ GeV
(dashed).}
\label{fig5}
\end{figure}

\begin{figure}
\caption{The maximally allowed $R_b$ (solid) as a function of $\tan\beta$ for
$m_{\chi_1}=60$ GeV and $m_{{\tilde t}_1} = \textstyle{1\over2}m_Z$.
We also plot the corresponding branching ratio $B(t\to bW)$ (dashed).
We have assumed $m_{{\tilde t}_2} = 250$ GeV and $\theta^{\tilde t}_{12}=1$.}
\label{fig6}
\end{figure}

\begin{figure}
\caption{Contours of maximally allowed values
$R_b = 0.2174$ (solid) and $0.2192$ (dotted) as well as
$B(t \to Wb)\ge 0.7$ (dashed) in the
$\mu-M_2$ plane where the heavy lines represent the LEP constraints and
$m_{\chi_1} > \textstyle{1\over2}m_Z$.
We assume $m_{{\tilde t}_1}=m_{{\tilde b}_1}=60$ GeV and
$m_{{\tilde t}_2}=m_{{\tilde b}_2}=250$ GeV. }
\label{fig7}
\end{figure}

\begin{figure}
\caption{The same as Fig.~7 but with the added constraint
$\theta^{\tilde t}=\theta^{\tilde b}$ so that $T=0$.\hspace{3cm}}
\label{fig8}
\end{figure}


\newpage
\input psfig
\centerline{\psfig{figure=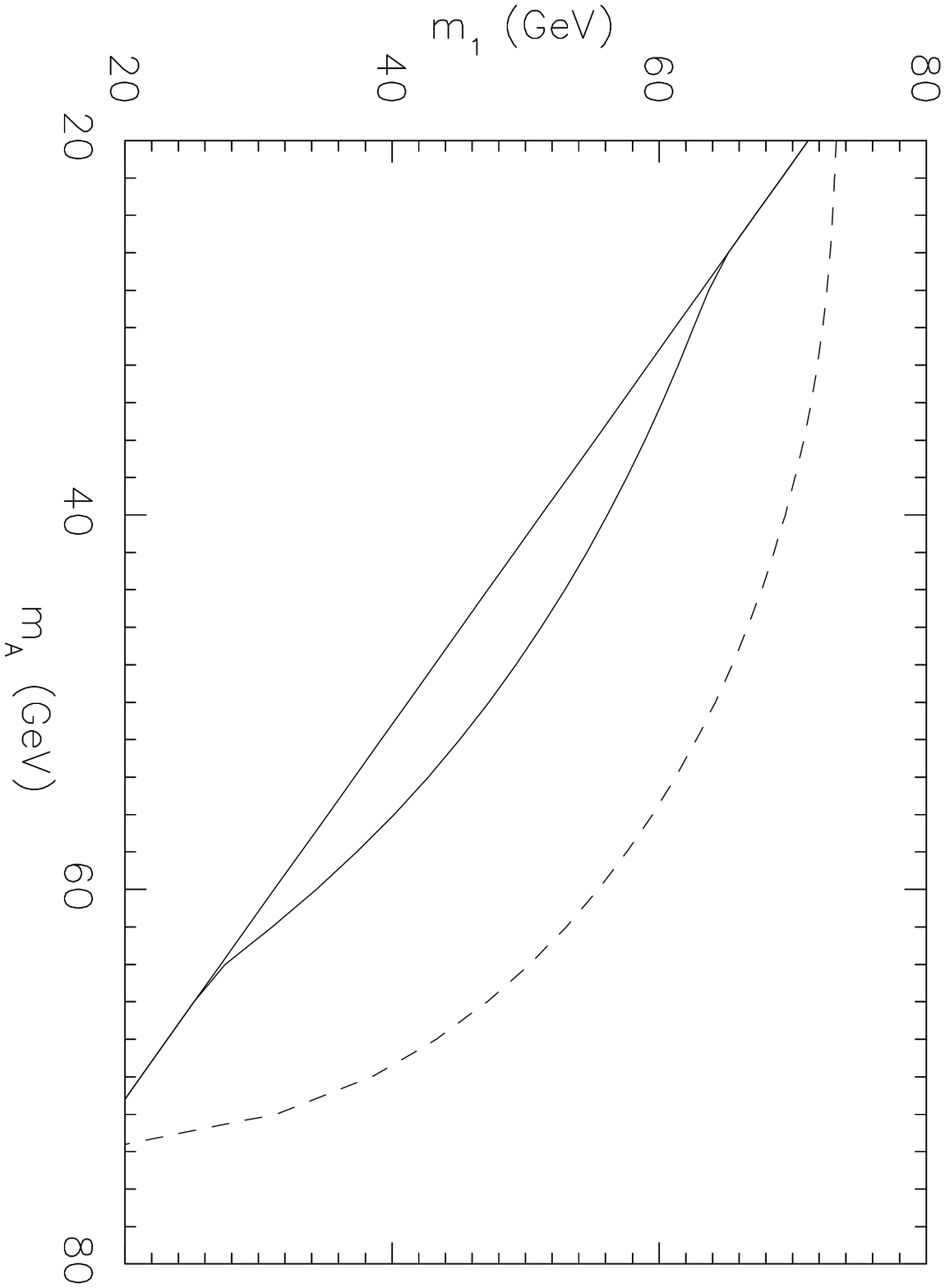,height=8.5in}}
\centerline{\bf Figure \ref{fig1} }

\newpage
\centerline{\psfig{figure=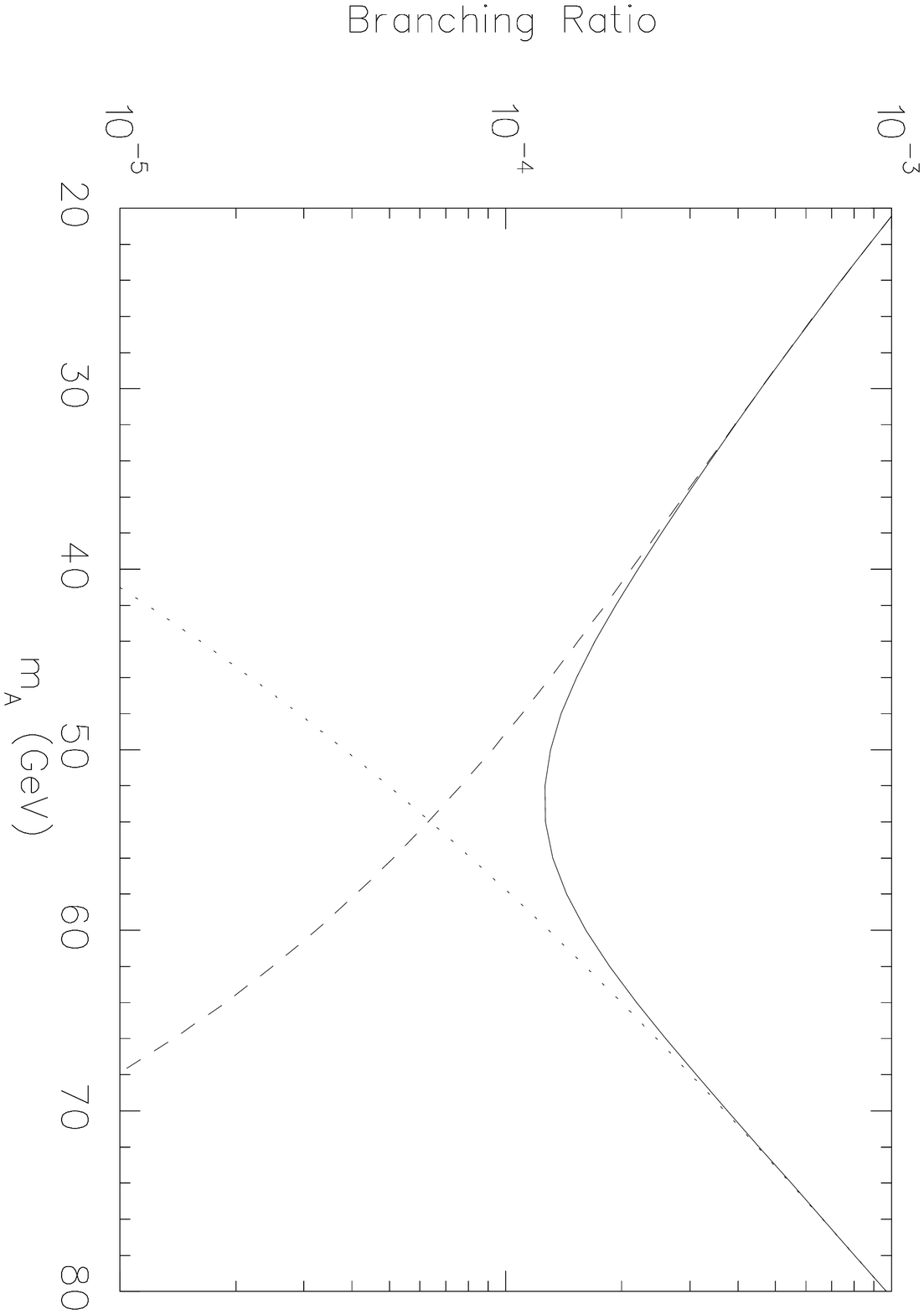,height=8.5in}}
\centerline{\bf Figure \ref{fig2} }

\newpage
\centerline{\psfig{figure=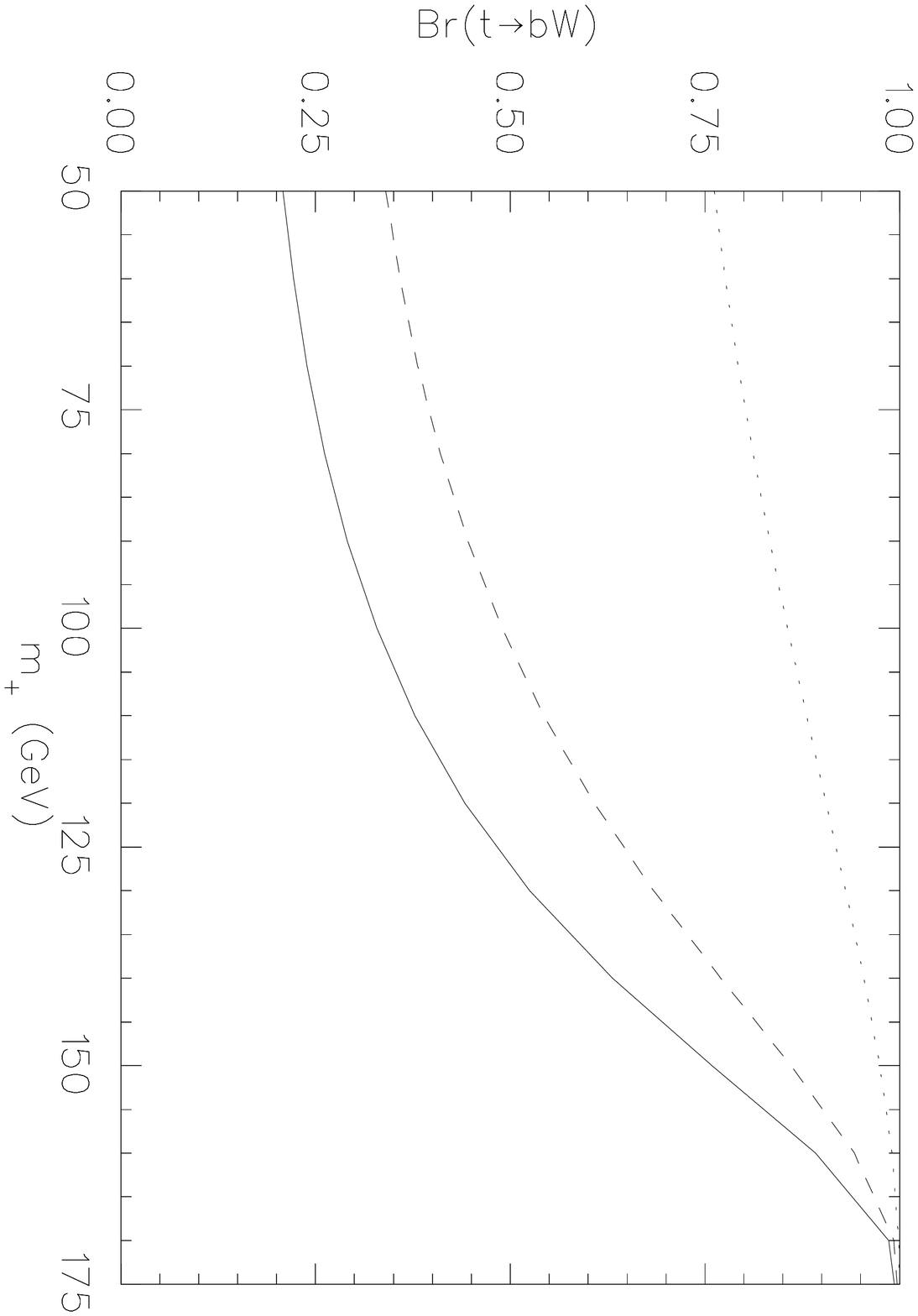,height=8.5in}}
\centerline{\bf Figure \ref{fig3} }

\newpage
\centerline{\psfig{figure=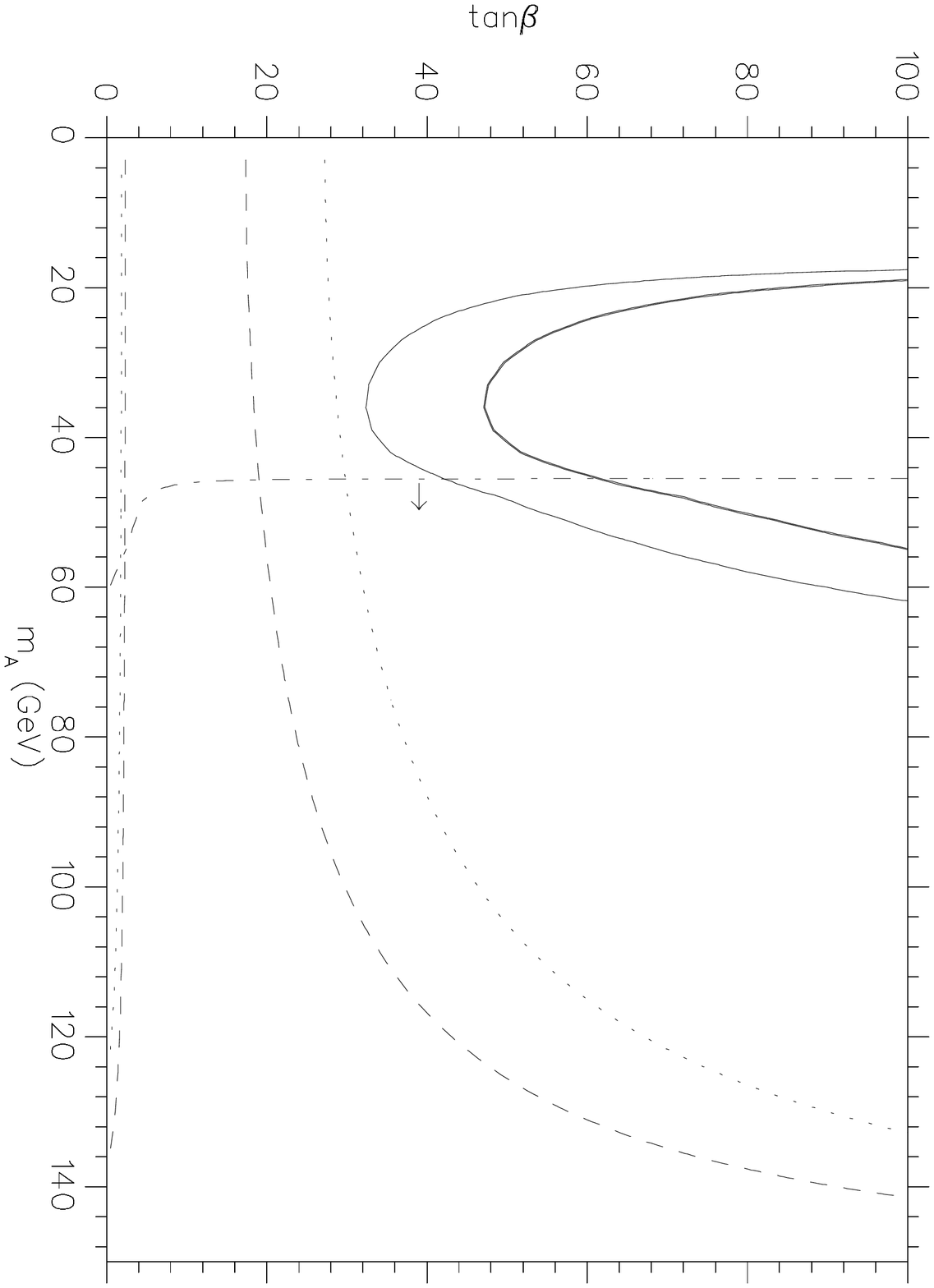,height=8.5in}}
\centerline{\bf Figure \ref{fig4} }

\newpage
\centerline{\psfig{figure=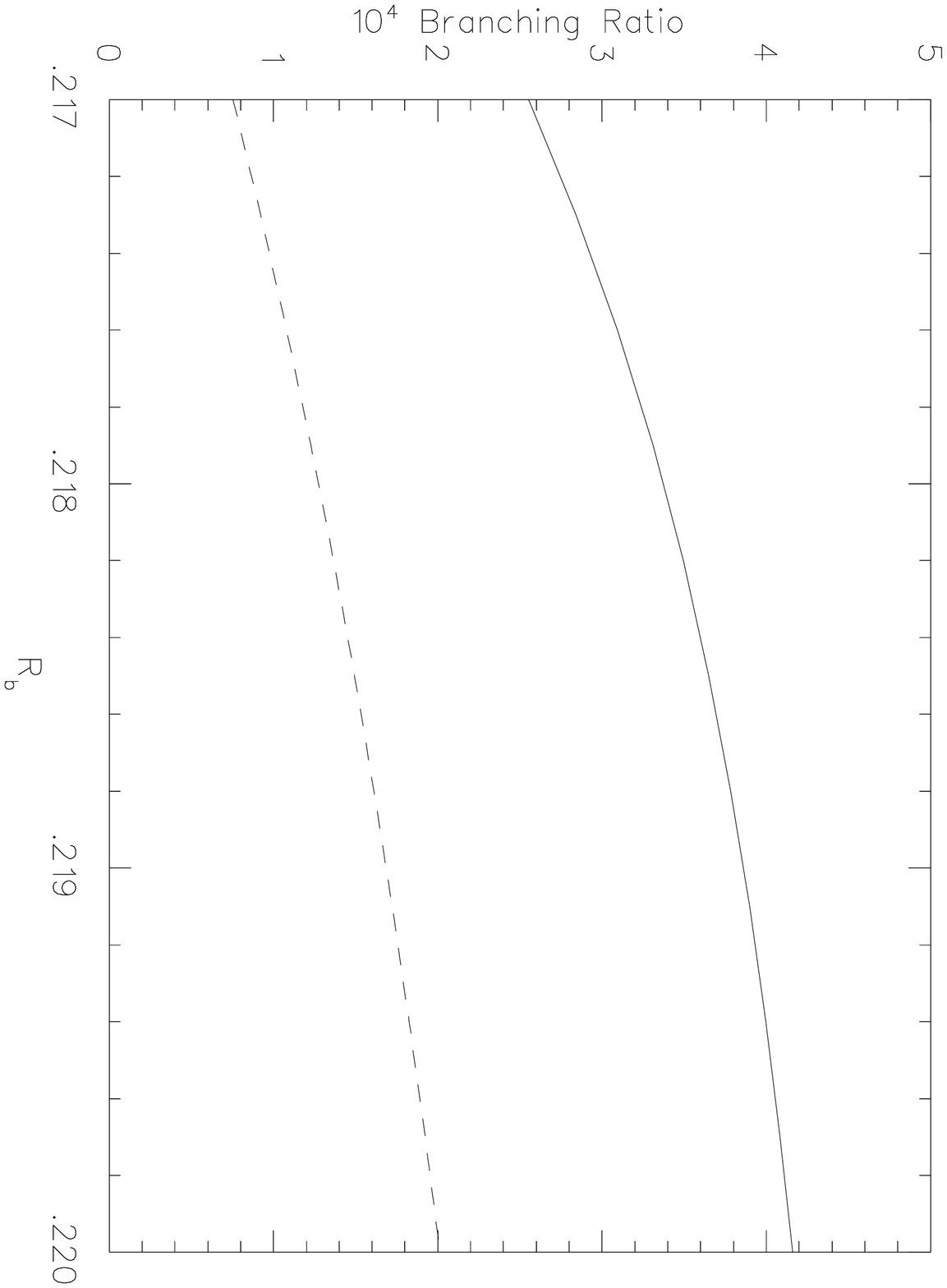,height=8.5in}}
\centerline{\bf Figure \ref{fig5} }

\newpage
\centerline{\psfig{figure=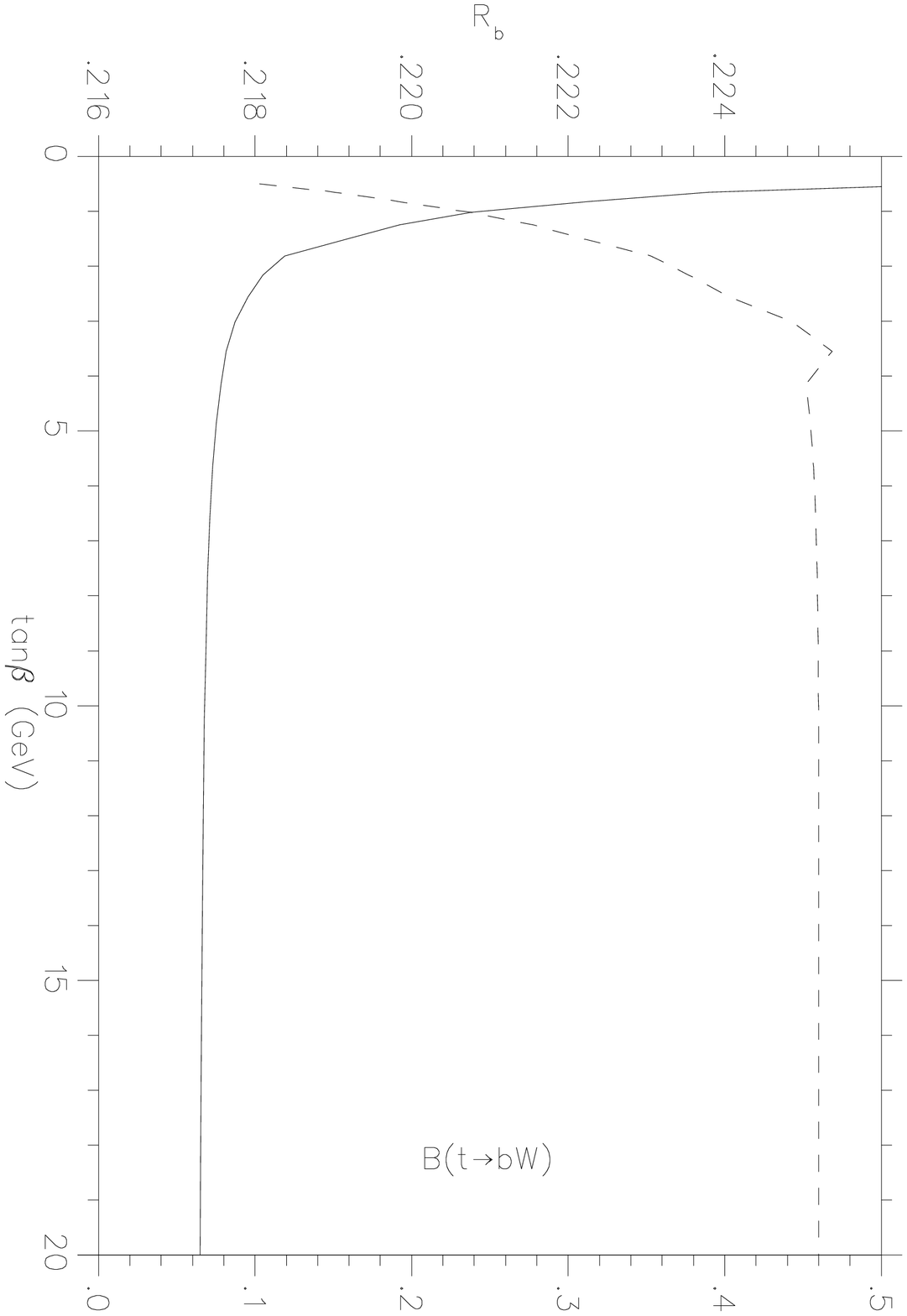,height=8.5in}}
\centerline{\bf Figure \ref{fig6} }

\newpage
\centerline{\psfig{figure=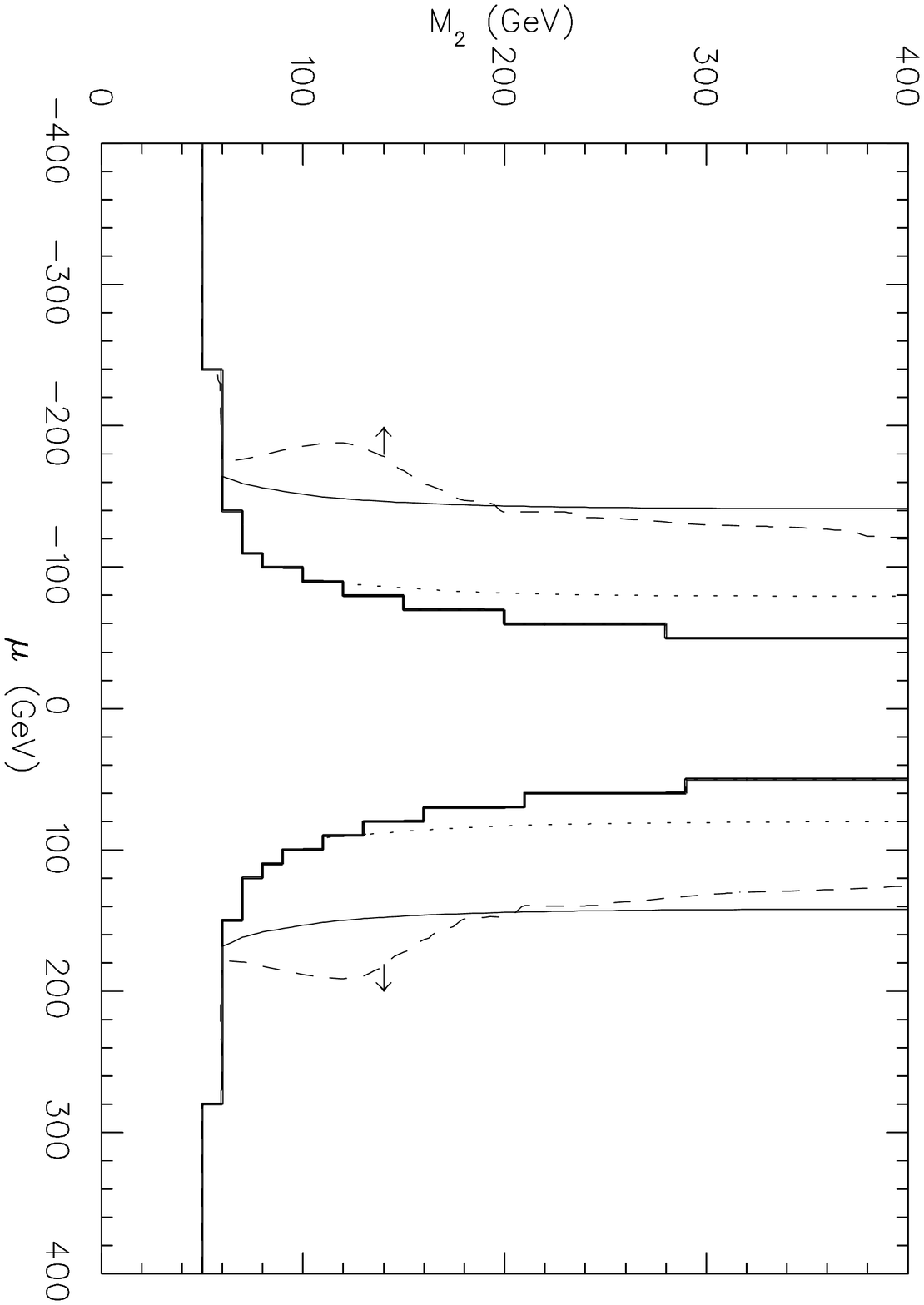,height=8.5in}}
\centerline{\bf Figure \ref{fig7} }

\newpage
\centerline{\psfig{figure=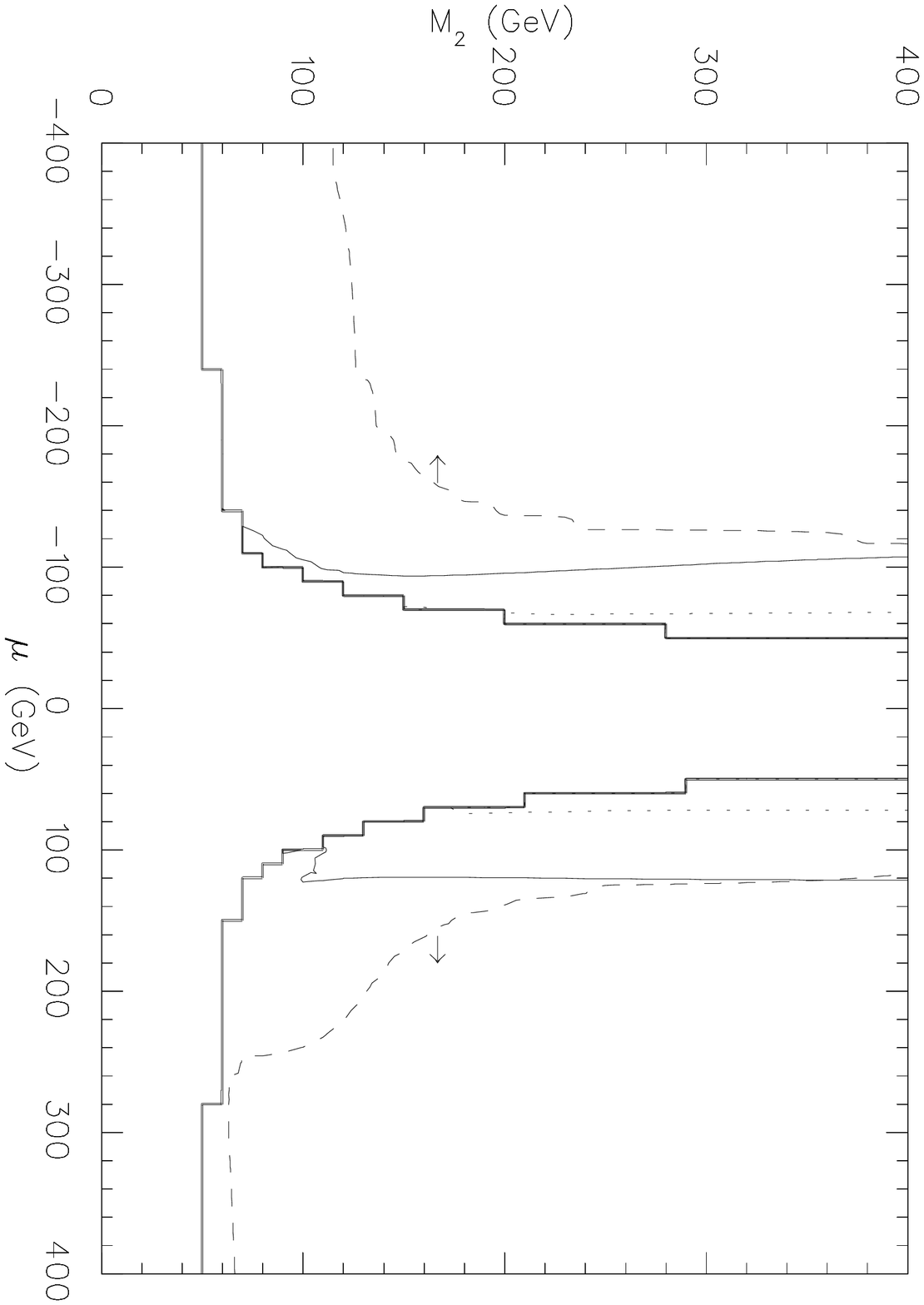,height=8.5in}}
\centerline{\bf Figure \ref{fig8} }
\end{document}